\documentclass[11pt]{article}
\usepackage{amssymb,amsfonts}
\usepackage[pdftex]{graphicx}
\DeclareGraphicsExtensions{.pdf}  \textheight24.5cm
\begin{document}

\renewcommand{\thefootnote}{\fnsymbol{footnote}}

\centerline{\Large \bf Note on Onsager's conjecture }
\vspace*{3mm} \centerline{Peter Stubbe\footnote[1]{retired from Max--Planck--Institut f\"ur Sonnensystemforschung, 37077 G\"ottingen, Germany. Contact by e-mail: peter-stubbe@t-online.de }} 
 
\renewcommand{\thefootnote}{\arabic{footnote}}

\vspace*{.5cm}{\small \begin{quote} Onsager conjectured that solutions of the incompressible Euler equations possessing a certain degree of roughness do not conserve the kinetic energy. Since, within the physical frame of Onsager's conjecture, the kinetic energy is the only occurring energy, and thus identical with the total energy, the implication would be that the conservation of energy is not absolute, but subject to the properties of mathematical solutions. Further, Onsager introduced the concept of anomalous dissipation of kinetic energy without viscosity. Both these aspects are critically discussed and their shortcomings unveiled. \end{quote} }

\vspace*{5mm}

{\bf 1}. In his highly regarded paper of 1949, Onsager [1] deals with questions concerning the development of incompressible flow into turbulent behaviour. He introduces the concept of energy cascades, transporting energy from larger to continually smaller scales, and he claims that a complete dissipation of all kinetic energy is possible thereby without the aid of viscosity.  As a side remark on the last page of his paper, Onsager states that turbulence cannot exist if the velocity field obeys a Lipschitz condition (he presumably means H\"older condition) with order $\theta$ greater than 1/3, and he adds that otherwise the energy is conserved.

This assertion has become known as Onsager's conjecture, and its standard formulation is as follows: The kinetic energy

\begin{equation} 
E_K = \frac{1}{2} N m \int {\bf v}^2 \, d^3{\bf r} 
\end{equation}

in a volume enclosed by a solid surface, or in entire space, is constant if ${\bf v}({\bf r},t)$ is H\"older continuous with exponent $\theta$ greater than 1/3, but decreases in time otherwise. In short

\begin{equation} 
\frac{dE_K}{dt} = 0 \mbox{\hspace*{5mm} if \hspace*{5mm}} \theta > \frac{1}{3}
\end{equation}
and  
\begin{equation} 
\frac{dE_K}{dt} < 0 \mbox{\hspace*{5mm} if \hspace*{5mm}} \theta \le \frac{1}{3}
\end{equation}

In (1) $N$ is the number density, $m$ the particle mass, {\bf v} the macroscopic velocity, and $d^3{\bf r}$ stands for $dx\,dy\,dz$.  Whereas (2) has been proved by Eyink [2] and Constantin, Weinan and Titi [3], (3) is still a conjecture..

The major assumption behind Onsager's conjecture is that the fluid under consideration is inviscid and incompressible, and the underlying equations are the so--called incompressible Euler equations 

\begin{equation} 
\frac{\partial \bf v}{\partial t} + ({\bf v} \cdot \nabla ) {\bf v} = -\nabla
\bar{p}
\end{equation}
\begin{equation} 
\nabla\cdot {\bf v} = 0
\end{equation}

where $\bar{p} = p / Nm$, and $p$ is the pressure.

It should be clarified what ``incompressible'' means. The condition (5) is not sufficient. It has to be assured that $N({\bf r},t)$ becomes a constant which, as the continuity equation 

\begin{equation} 
\frac{\partial N}{\partial t} + {\bf v} \cdot \nabla N  = -\, N \, \nabla \cdot {\bf v}
\end{equation}

shows, is achieved by (5) and  

\begin{equation} 
\nabla N = 0
\end{equation} 

Both (5) and (7) are needed to secure the validity of (4) and (1). Otherwise, $N$ would stand in front of the Nabla--operator on the right--hand side of (4) and behind the integral sign in (1), with the consequence that the continuity equation (6) would have to become a member of the system of equations.

\vspace*{3mm}

{\bf 2}. Onsager's conjecture has remained a subject of considerable mathematical interest (e.g., [4]--[6], and references cited therein), with (3) as still existing challenge. The question is whether this mathematical work has its meaning only within its mathematical environment, or beyond for physics, too.
 
This question is connected with the question of the physical substance of Onsager's conjecture. If one had a physical system with at least one other form of energy in addition to the kinetic energy, then the conjecture would be of little interest because conversions between different forms of energy are a normality. If, on the other hand, the kinetic energy would be the only energy in the system, and thus identical with the total energy, then the non-conservation of kinetic energy expressed by (3) would have a sensational character since it would disobey the most fundamental law of physics, the conservation of total energy in a closed system.

\vspace*{3mm}

{\bf 3}. A look at the starting equations (4) and (5) shows that there is no possiblity to derive any other energy equation than one for the kinetic energy (c.f.~point 5 below), and so the physical content of Onsager's conjecture (3) is indeed that the total energy is not conserved for specific mathematical solutions. Since it is not seriously thinkable that the existence of certain special solutions could overrule a fundamental physical law, the foundation of Onsager's conjecture should be critically examined:

If one derives the Euler equation in a heuristic way (Newton's second law, applied to a pressure gradient force), one may be led to the impression that anything seemingly plausible can be used as a closing equation. In the given case this is the incompressibility condition (5). Common as this may be, it is invalid nonetheless. 

The point is that any transport equation of hydrodynamics is part of an infinite system of equations (e.g., [7]--[9], particularly eq.~(9) of [9]), and it requires very restrictive truncation conditions to extract a finite system of equations from the infinite system. 

To lowest level of approximation, a closed set of equations is obtained for the three fluid variables $N$, {\bf v} and temperature $T
$, given by the continuity equation (6) above and, e.g., eqs.~(24) and (25) of [9]: 

\begin{equation} 
\frac{\partial \bf v}{\partial t}  + ({\bf v} \cdot \nabla ){\bf v} =  - \frac{1}{Nm} \nabla p = - \, V^2 \left[\, \frac{\nabla T}{T} + \frac{\nabla N}{N} \, \right] 
\end{equation}
\begin{equation} 
\frac{\partial T}{\partial t} + {\bf v} \cdot \nabla T = - \frac{2}{3} \, T \, \nabla \cdot {\bf v}
\end{equation}

where $V$ is the thermal velocity ($V^2 = KT/m$ , $K$ = Boltzmann's constant) and $p = NKT$. The corresponding Navier--Stokes system is given by eqs.~(42) and (43) of [9].

The equation system [(6),(8),(9)] is complete. The number of equations matches the number of unknowns, and hence there is no room for any other equation. In particular, eq.~(5), $\nabla \cdot {\bf v} = 0$, cannot be used to supersede the thermal equation (9). If this is done nonetheless, it has to be shown a posteriori that the solutions satisfy (9). This equation does not lose its existence only because incompressibility has been assumed.

The permissible way to employ the incompressibility conditions (5) and (7) is to simplify (6), (8) and (9) to obtain $N$ = const.~and

\begin{equation} 
\frac{\partial \bf v}{\partial t}  + ({\bf v} \cdot \nabla ){\bf v} =   - \, V^2 \, \frac{\nabla T}{T}  = - \, \frac{K}{m} \, \nabla T
\end{equation}
\begin{equation} 
\frac{\partial T}{\partial t} + {\bf v} \cdot \nabla T =  0
\end{equation}

The equations [(10),(11)] have to be used in place of [(4),(5)] if the incompressibility conditions (5) and (7) are applied in a proper way.

\vspace*{3mm}

{\bf 4}. When one compares the full Euler system [(6),(8),(9)] with the so--called incompressible Euler equations [(4),(5)], it appears difficult to understand how the latter could ever have reached their enormous popularity (e.g., see the review article [10]). Should it not be obvious to ask what the pressure gradient term $\nabla p$ means when there exists no density gradient $\nabla N$?  Naturally it means that the whole burden of driving the velocity field is placed on a temperature gradient. But how then should it be possible to expect anything of physical significance if no equation for the temperature is formulated? And why, at all, should it be more plausible to ignore $\nabla N$ than $\nabla T$? Would it, in the end, not be most plausible to ignore both to reach the ultimate form of simplicity, the incompressible isothermal equation 

\begin{equation} 
\frac{\partial \bf v}{\partial t} + ({\bf v} \cdot \nabla ) {\bf v} = 0 \mbox{\hspace*{1mm} ?}
\end{equation}

\vspace*{3mm}

{\bf 5}. The kinetic energy density of an ensemble of particles with distributed velocities {\bf u} is given by

\begin{equation} 
w = \frac{1}{2} \, m \int f \, {\bf u}^2 d^3 {\bf u} = \frac{1}{2} \, m \int f \, {\bf v}^2 d^3 {\bf u} + \frac{1}{2} \, m \int f \,
( {\bf u} - {\bf v} )^2 d^3 {\bf u}
\end{equation}

where $f({\bf u}, {\bf r}, t)$ is the distribution function in phase space ($ \int f \, d^3 {\bf u} = N $, $ \int f {\bf u} \, d^3 {\bf u} = N{\bf v}) $. We define

\begin{equation} 
w_K = \frac{1}{2} \, m \int f \, {\bf v}^2 d^3 {\bf u} = \frac{1}{2} Nm{\bf v}^2
\end{equation}
\begin{equation} 
w_I = \frac{1}{2} \, m \int f \, ( {\bf u} - {\bf v} )^2 d^3 {\bf u} = \frac{3}{2} NKT = \frac{3}{2}\, p
\end{equation}
 
where $w_K$ is the kinetic energy density due to the bulk motion and $w_I$ the internal energy density due to motion about the average.  From (8), multiplied by {\bf v}, and (9), the energy equations

\begin{equation} 
\frac{\partial w_K}{\partial t} + \nabla \cdot ( w_K \, {\bf v}) = - \, {\bf v}\cdot \nabla p 
\end{equation}

\begin{equation} 
\frac{\partial w_I}{\partial t} + \nabla \cdot (w_I \, {\bf v}) = - \, p \, \nabla \cdot {\bf v} 
\end{equation}

are obtained, and these add up to

\begin{equation} 
\frac{\partial w}{\partial t} + \nabla \cdot (w \, {\bf v}) = - \, \nabla \cdot(p \,{\bf v}) 
\end{equation}

Eq.~(18) shows that the energy $E = \int w \, d^3 {\bf r}$ inside a closed solid surface is constant, whereas according to (16) and (17) the individual energies $E_K = \int w_K \, d^3 {\bf r}$ and $E_I = \int w_I \, d^3 {\bf r}$ can change in arbitrary ways, provided their sum remains constant. If derived on the basis [(4),(5)], the energy equations (17) and (18) would be absent and the kinetic energy equation (16) the only one. 

In the case $\nabla \cdot {\bf v} = 0$ the right--hand side term in (17) vanishes, and the right--hand side term in (16) can be altered into $ \nabla \cdot(p \,{\bf v}) $, with the consequence that now all three energies $E_K$, $E_I$ and $E$ are constant. In particular, it has been shown thereby that 

\begin{equation} 
\frac{dE_K}{dt} = 0 \mbox{\hspace*{5mm} if \hspace*{5mm}} \nabla \cdot {\bf v} = 0 
\end{equation}

This statement holds irrespective of the particular mathematical properties of ${\bf v} ({\bf r},t)$, and it disproves Onsager's conjecture (3). One may use (19) as a filter: Mathematical solutions based on the Euler system that would violate (19) would be physically invalid.
 
\vspace*{3mm}

{\bf 6}. While it is true that kinetic energy can migrate from larger to smaller structures, it is also true that this does not change the character of the energy, kinetic energy remains kinetic energy. Therefore, the concept of anomalous dissipation introduced by Onsager [1] (dissipation without viscosity) cannot be viable unless a mechanism is given that would irreversibly convert kinetic energy into another energy form. 

Such a mechanism is not available if one remains within the frame of either equations [(4),(5)], or [(8),(9)], or [(10),(11)]. As the energy equations (16) and (17) show, there exists no term that would irreversibly transfer kinetic energy into another form of energy. The energy exchange term in (16), $ \dot{w}_{K\leftrightarrow P} \equiv - \, {\bf v} \cdot \nabla p $, describes the mutual conversion of kinetic and potential energy of an ensemble of particles moving along or against the pressure gradient, and in (17) $ \dot{w}_{I\leftrightarrow P} \equiv - \, p \, \nabla \cdot {\bf v} $ describes the mutual conversion of internal and potential energy due to compression or dilatation. 

\vspace*{3mm}

{\bf 7}. General energy equations without limiting assumptions are given by

\begin{equation} 
\frac{\partial w_K}{\partial t} + \nabla \cdot ( w_K \, {\bf v}) = - \, {\bf v}\cdot \left[\, \nabla p + \nabla \cdot (\textsf{p})^{\rm o} \, \right]
\end{equation}
\begin{equation} 
\frac{\partial w_I}{\partial t} + \nabla \cdot (w_I \, {\bf v}) = - \, \left[\, p \, \nabla \cdot {\bf v} +(\textsf{p})^{\rm o} : \nabla {\bf v} \right]  - \nabla \cdot {\bf q}
\end{equation}
\begin{equation} 
\frac{\partial w}{\partial t} + \nabla \cdot (w \, {\bf v}) = - \nabla \cdot (p \, {\bf v}) -  \nabla\cdot [ (\textsf{p})^{\rm o}\cdot {\bf v}]  - \nabla \cdot {\bf q}
\end{equation}

(see eqs.~(46)--(48) of [9]). Here $(\textsf{p})^{\rm o}$ is the traceless part of the pressure tensor and {\bf q} the heat flux vector. Writing out the terms on the right--hand sides of (20) and (21) in component form and defining

\begin{equation} 
\dot{w}_{K\leftrightarrow P} = - \, {\bf v}\cdot \nabla p - v_x
\frac{\partial}{\partial x} (p_{xx} - p) - v_y \frac{\partial}{\partial y}
(p_{yy} - p) - v_z \frac{\partial}{\partial z} (p_{zz} - p)
\end{equation}
\begin{equation} 
\dot{w}_{K\leftrightarrow K} = - \, \frac{\partial}{\partial x} \left( v_y
p_{xy} + v_zp_{xz}\right) - \, \frac{\partial}{\partial y} \left( v_zp_{yz} +
v_x p_{xy} \right) - \, \frac{\partial}{\partial z} \left( v_x p_{xz} +
v_yp_{yz}\right)
\end{equation}
\begin{equation} 
\dot{w}_{K\rightarrow I} = - \, p_{xy}\left(\frac{\partial v_x}{\partial y} +
\frac{\partial v_y}{\partial x} \right) - \, p_{yz}\left(\frac{\partial
v_y}{\partial z} + \frac{\partial v_z}{\partial y} \right) - \,
p_{xz}\left(\frac{\partial v_z}{\partial x} + \frac{\partial v_x}{\partial z}
\right)
\end{equation}
\begin{equation} 
\dot{w}_{I\leftrightarrow P} = - \, p \, \nabla \cdot {\bf v} - (p_{xx}-p)\frac{\partial v_x}{\partial x} - \, (p_{yy}-p)\frac{\partial
v_y}{\partial y} - \, (p_{zz}-p)\frac{\partial v_z}{\partial z}
\end{equation}
\begin{equation} 
\dot{w}_{I\leftrightarrow I} = - \nabla \cdot {\bf q}
\end{equation}

the energy equations can be written in the transparent form

\begin{equation} 
\frac{\partial w_K}{\partial t} + \nabla \cdot ( w_K
 \, {\bf v})  =  \dot{w}_{K\leftrightarrow P} + \dot{w}_{K\leftrightarrow K} - \dot{w}_{K\rightarrow I}
\end{equation}
\begin{equation} 
\frac{\partial w_I}{\partial t} + \nabla \cdot (w_I \, {\bf v}) =
\dot{w}_{I\leftrightarrow P} + \dot{w}_{I\leftrightarrow I} +
\dot{w}_{K\rightarrow I}
\end{equation}

The physical meaning of $\dot{w}_{K\leftrightarrow P}$ and $\dot{w}_{I\leftrightarrow P}$ is as before. The terms $\dot{w}_{K\leftrightarrow K}$ and $\dot{w}_{I\leftrightarrow I}$ express the spatial redistribution of kinetic and internal energy, respectively, and the mathematical form of (24) and (27) shows that this redistribution does not change the available energy, but only its local distribution. In contrast, $\dot{w}_{K\rightarrow I}$ describes the irreversible conversion of kinetic into internal energy. It will be seen below that $\dot{w}_{K\rightarrow I}$ is an unconditionally positive quantity, as it should. 

The dual process of cascading with final loss of kinetic energy is contained in the combined action of the terms $\dot{w}_{K\leftrightarrow K}$ and $\dot{w}_{K\rightarrow I}$, and eq.~(29) shows that the kinetic energy lost in (28) is fully transferred into the internal energy reservoir so that energy conservation is secured. 

Up to this point $(\textsf{p})^{\rm o}$ and {\bf q} are undetermined. Expressions pertaining to the Navier--Stokes approximation are, e.g., given by eqs.~(37) and (39) of [9], and this completes the derivation. Using the expression for $(\textsf{p})^{\rm o}$ in (24) and (25), $\dot{w}_{K\leftrightarrow K}$ and $\dot{w}_{K\rightarrow I}$ are obtaind as 

\begin{equation} 
\dot{w}_{K\leftrightarrow K} = \eta \left\{ \frac{\partial}{\partial x} \left[ v_y \left( \frac{\partial v_x}{\partial y} + 
\frac{\partial v_y}{\partial x} \right) + v_z \left( \frac{\partial v_x}{\partial z} + \frac{\partial v_z}{\partial x} \right) \right] 
+ ..... + ..... \, \, \right\}
\end{equation}
\begin{equation} 
\dot{w}_{K\rightarrow I} =  \eta \, \bigg[ \left( \frac{\partial v_x}{ \partial y} 
+ \frac{\partial v_y}{\partial x} \right)^{\mbox{\hspace*{-1.2mm}}2} + \left( \frac{\partial v_y}{\partial z} + \frac{\partial v_z}{\partial y} \right)^{\mbox{\hspace*{-1.2mm}}2} +\left( \frac{\partial v_z}{\partial x} + \frac{\partial v_x}{\partial z} \right)^{\mbox{\hspace*{-1.2mm}}2} \, \bigg]
\end{equation}

Here $ \eta $ is the dynamic viscosity. Eq.~(31) shows that indeed $\dot{w}_{K\rightarrow I}$ is an entirely positive quantity. 
 
\vspace*{3mm}
 
{\bf 8}. Looking at eq.~(30), one may wonder why viscosity should be needed for a restructuring of scales. Actually, it is not really viscosity that matters, but the existence of a fully developed pressure tensor (which is one and the same within the Navier--Stokes approximation). The underlying assumption in the Euler approximation is that the system is completely isotropized, i.e, the three diagonal elements in the pressure tensor are equal, and the elements outside the diagonal zero ($p_{xx} = p$, $p_{xy} = 0$, and correspondingly for the other elements), and furthermore ${\bf q} = 0$. The equations (23)--(27) show that almost the entire physics is extinguished by these restrictions, and so the Euler system is seen to be of very limited value, even in its complete form [(6),(8),(9)].

In the Navier--Stokes approximation, all elements in the pressure tensor are occupied, however subject to the limiting conditions $|p_{xx} - p| \ll p$ and $|p_{xy}| \ll p $, etc.~(see [9] for an extensive discussion). As a consequence, only small deviations from equilibrium will be physically allowed. Otherwise, no transport equation will be available, and the equation to be used will be the kinetic equation for the distribution function in 6--dimensional phase space,  $f({\bf u}, {\bf r}, t)$. Another limiting aspect is the condition that the spatial scale of the process must be larger than the mean free path, and the time scale longer than the time between collisions. Under regular circumstances, however, this is not a serious restriction.   
 
\vspace*{3mm}

{\bf 9}. To summarize: (a) It is not physically justified to close Euler's equation (and likewise the Navier--Stokes equation) by the incompressibility condition $\nabla \cdot {\bf v} = 0$. (b) Onsager's conjecture (3) expresses a physical impossibility, and a physically valid proof will thus not be possible. (c) The idea that endless cascading from larger to smaller scales could explain a loss of kinetic energy without the aid of viscosity is erroneous. The total kinetic energy is not changed by a restructuring. An irreversible loss occurs by conversion into internal energy, and this requires viscosity. (d) As a matter of principle, questions regarding the irreversible loss of kinetic energy are outside the range of the Euler approximation. (e) From a mathematics point of view, eqs.~[(4),(5)] are to be regarded as PDEs without relation to physical reality. 

\vspace*{3mm}

\vspace*{1 cm} 
\centerline {\bf References}
\vspace*{3mm}

[1] L.~Onsager, {\it Statistical hydrodynamics}, Nuovo Cimento Suppl.~{\bf 6} (1949), 279--287.

[2] G.~Eyink, {\it Energy dissipation without viscosity in ideal hydrodynamics I. Fourier analysis and local energy transfer}, Physica D {\bf 78} (1994), 222--240.

[3] P.~Constantin, E.~Weinan and E.~Titi, {\it Onsager's conjecture on the
energy conservation for solutiuons of Euler's equation}, Comm.~Math.~Phys.~{\bf
165} (1994), 207--209.

[4] A.~Cheskidov, P.~Constantin, S.~Friedlander and R.~Shvydkoy, {\it Energy
conservation and Onsager's conjecture for the Euler equations}, Nonlinearity
{\bf 21} (2008), 1233--1252.

[5] C.~De Lellis and L.~Szekelyhidi, {\it Dissipative continuous Euler flows},
Inventiones Mathematicae {\bf 193}, Issue 2 (2013), 377--407.

[6] T.~Buckmaster, C.~De Lellis and L.~Szekelyhidi, {\it Dissipative Euler flows with Onsager--critical spatial regularity}, arXiv:1404.6915 (2014)

[7] S.~Chapman and T.~Cowling, {\it The Mathematical Theory of Non--Uniform
Gases}, Cambridge University Press (1970).

[8] R.~Schunk, {\it Mathematical structures of transport equations for multispecies flows}, Rev.~Geophys.~Space Phys.~{\bf 15} (1977), 429--445.

[9] P.~Stubbe, {\it The Euler and Navier--Stokes equations revisited}, arXiv:1506.04561 (2015).

[10] P.~Constantin, {\it On the Euler equation of incompressible fluids}, Bull.~Amer.~Math.~Soc.~{\bf 44} (2007), 603--621.

\end{document}